**Graph Neural Network Force Fields (GPTFF-mol) for Organic Molecules from Optimization Trajectories (OpenGEM26)**

*Yifan Huang,[1,2] Fankai Xie,[1,2] Jiangnan Zheng,[1,2] Tenglong Lu,[1,3] Sheng Meng, [1*] Miao Liu, [1,3*]*

[1] Beijing National Laboratory for Condensed Matter Physics, Institute of Physics, Chinese Academy of Sciences, Beijing 100190, China;

[2] University of Chinese Academy of Sciences, Beijing 100049, China

[3] Dongguan Institute of Materials Science and Technology, Dongguan, Guangdong 523808, China

* E-mail: smeng@iphy.ac.cn, mliu@iphy.ac.cn

**Abstract**

Density functional theory (DFT) serves as a reliable tool for atomistic molecular simulations, while machine learning potentials have become powerful complements to balance accuracy and efficiency. In this work, we release OpenGEM26 (Open Generated Ensemble of Molecules, 2026), a large-scale dataset comprising 200,000 unique molecules and 4.4 million conformations composed of H, C, N, O, S and Cl with up to ten heavy atoms. All calculations are carried out at the ωB97X-D/Def2-SVP and Def2-TZVP levels with dispersion corrections, and complete structural optimization trajectories and abundant non-equilibrium structures are recorded. Statistical analyses confirm that this dataset covers a broader conformational space than QM9 in terms of energy, bond lengths and bond angles. A graph neural network-based potential GPTFF-mol is trained using the new dataset, achieving an energy mean absolute error of 16 meV/molecule, which is equivalent to 0.82meV/atom, and superior force prediction performance compared with ANI-2x. Validated by butane rotation and keto-enol tautomerization tests, the model accurately describes molecular dynamical behaviors and reaction barriers at distorted geometries. This work provides a high-quality resource and robust ML potential for efficient simulations of sulfur- and chlorine-containing organic molecules.

**INTRODUCTION**

The accurate description of interactions between atoms and molecules is the cornerstone of reliable simulations in physics, chemistry, biology and materials science. Quantum mechanical simulation methods, including Kohn–Sham density functional theory (KS-DFT, hereafter abbreviated as DFT) and quantum chemical calculations, have become the most authoritative and accurate approaches for atomistic-scale computational research. These techniques have been widely applied to calculate properties of diverse systems, such as battery materials[1–3], catalysis[4–6] and superconductivity[7,8]. With the rapid development of quantum mechanical simulations, such methods are frequently combined with molecular dynamics or classical force field approaches to describe chemical reactivity in large-scale systems.

Despite advances in computational hardware, conventional quantum mechanical algorithms exhibit a computational complexity scaling of $O(N^3)$ with the total number of electrons N. This characteristic severely limits the applicability of such methods to large-scale systems. To address this bottleneck, machine learning (ML)-based interatomic potentials have been developed[9–13]. Such models feature low computational complexity while maintaining high prediction accuracy. Superior performance relative to traditional force fields is achieved without introducing excessive computational overhead, making ML potentials a mainstream research tool in modern computational science[14–18]. In principle, neural networks serve as a mathematical framework capable of fitting arbitrary smooth continuous functions, provided that sufficient parameters and training data are available. Classical force fields typically contain only dozens of parameters, whereas contemporary ML-based force fields are constructed with up to millions of parameters. Accordingly, ML potentials deliver substantially higher performance and accuracy compared with conventional classical force fields.

The inherent flexibility of neural network architectures enables efficient extraction of complex information from reference datasets. Meanwhile, this flexibility imposes strict requirements on large-scale, high-fidelity training sets. Models trained on incomplete or poorly curated datasets are prone to underfitting or overfitting, and robust generalization performance cannot be guaranteed. For this reason, the construction of comprehensive, high-quality datasets is identified as an essential prerequisite for the reliable training of neural network-based interatomic potentials.

The QM9 Chemistry dataset[19], comprising approximately 133,885 small organic molecules with up to nine heavy atoms (C, O, N, F) and their DFT-computed properties at the B3LYP/6-31G(2df,p) level, has served as the foundational benchmark for developing next-generation machine learning potentials and force fields. Smith, Isayev, and Roitberg pioneered this field with ANI-1[20], a dataset beyond equilibrium geometries by sampling about 20 million off-equilibrium conformations. Using this dataset, they trained a transferable deep neural network potential based on modified Behler–Parrinello symmetry functions[21]. ANI-1x used active learning to identify poorly predicted configurations, reducing the dataset to about 5 million conformations, and ANI-1ccx, which applied transfer learning from DFT to high-level CCSD(T)/CBS data to approach coupled-cluster accuracy with only 10% expensive post-Hartree-Fock calculations[22]. In parallel, Schütt et al. developed SchNet[23,24], a continuous-filter convolutional neural network specifically designed for molecules and materials that achieves 0.013 eV MAE for energy on QM9. Unke and Meuwly's PhysNet[25] introduced explicit electrostatics through neural-network-predicted partial charges, achieving about 0.01 eV energy errors while simultaneously predicting energies, forces, dipole moments, and charges via reverse-mode automatic differentiation for spectroscopic and long-range interaction applications. Complementing these neural network approaches, Chmiela, Tkatchenko, and Müller developed GDML[26] and its symmetric extension sGDML[27], kernel-based methods that directly predict atomic forces with guaranteed energy conservation and automatic recovery of permutation symmetries, demonstrating competitive accuracy with enhanced data efficiency for small training sets. The PiNN[28] achieved 0.012 eV MAE for QM9 internal energies and 0.018 Debye for dipole moments through modular atomic neural network architectures, while Faber et al.'s FCHL[29] representation combined with Gaussian process regression attained 0.022 eV MAE for atomization energies. More recently, equivariant deep learning potentials such as Allegro have extended these capabilities through E(3)-equivariant message passing and multi-objective optimization across diverse datasets, enabling unified potentials spanning from small QM9 molecules to larger systems[30]. Collectively, these developments have established QM9-trained machine learning potentials as transformative tools for molecular dynamics simulations, reaction pathway exploration, and high-throughput screening of organic molecules, fundamentally bridging the accuracy-efficiency trade-off that has historically constrained computational chemistry applications.

Besides the widely-used QM9 dataset, numerous complementary datasets and projects have been developed to advance machine learning potentials and force fields for organic molecules and beyond. The GDB family of enumerated databases—GDB-11 (26 million molecules), GDB-13 (970 million molecules), and GDB-17 (over 166 billion molecules)—systematically explores the full chemical compound space of drug-like small organic molecules with increasing size limits[31–33]. The quantum-mechanical benchmark series progressed from QM7 (7,165 molecules with PBE0 energies) through QM7b (with 13 additional electronic properties) and QM7-X (extensive non-equilibrium coverage) to QM8 (~20,000 structures with excited-state TDDFT and CC2 properties), each expanding the diversity of computable molecular properties[34–37]. For training transferable neural network potentials, the ANI family of datasets—ANI-1 (20 million non-equilibrium conformations), ANI-1x (5 million calculations with density-derived properties and active learning), and ANI-1ccx (500,000 CCSD(T) energies for coupled-cluster accuracy)—provides systematically improved training data through transfer and active learning strategies[20,22]. Molecular dynamics-focused benchmarks include MD17 (ab initio trajectories for eight organic molecules at 500K), its corrected revision rMD17 with tighter SCF convergence, and ISO17 (129 $C_7O_2H_{10}$ isomers with 5,000 conformations each), all providing critical energy-force pairs for dynamics-capable potentials[26,38,39]. Higher-accuracy extensions such as QM9-G4MP2 and G4MP2-heavy enable Δ-learning from DFT to coupled-cluster accuracy and extrapolation to larger systems[40], while the Alchemy dataset[41] (119,487 molecules up to 14 heavy atoms) and AGZ7[42] (140,000 molecules covering 13 elements through systematic fragmentation) expand elemental diversity and size range. Large-scale spectroscopic and structural databases including OE62[43] (61,489 molecules with orbital eigenvalues), PubChemQC[44] (millions of DFT-computed molecules from PubChem), and the specialized tmQM[45] (86,000 transition metal complexes) further broaden applicability to diverse chemical systems. More recent innovations include QH9[46] (130,831 Hamiltonian matrices for direct quantum operator learning), COMP6[47] (composite benchmark for consistent evaluation), Rad-6[48] (radical species for open-shell challenges), and MB08-165[49] (artificial molecules for controlled benchmarking), collectively establishing a rich ecosystem of datasets that enable development of increasingly accurate, generalizable, and efficient machine learning potentials across chemical space.

Despite the abundant existing datasets for ML interatomic potential training, several critical limitations remain unresolved. Most mainstream benchmark databases including QM9 adopt the B3LYP/6-31G(2df,p) DFT computational level, which limits the overall accuracy of reference data. Furthermore, the molecular compositions of these datasets are predominantly restricted to C, H, N, O and F. Even datasets with relatively high computational accuracy generally lack coverage of sulfur and chlorine, two elements widely present in organic compounds. In addition, most datasets with many kinds of elements do not have sufficient non-equilibrium conformations and complete structural optimization trajectories. The performance of corresponding models is therefore restricted when describing molecular dynamical behaviors and chemical reactions.

To fill the above research gaps, we construct a new dataset, OpenGEM26 (Open Generated Ensemble of Molecules, 2026), derived from the GDB-13 database. The dataset contains approximately 200,000 distinct molecules and 4.4 million molecular conformations, with molecules consisting of H, C, N, O, S and Cl and a maximum of ten heavy atoms. All quantum chemical calculations are performed at the higher-precision ωB97X-D/Def2-SVP and Def2-TZVP level[50–52], which incorporates empirical atom–atom dispersion corrections to ensure reliable reference values. This dataset also records full structural optimization trajectories and includes abundant non-equilibrium structures, enabling far broader coverage of chemical and conformational space. Such features make the dataset well-suited for training robust, high-accuracy machine learning molecular potentials.

## DATASET

**Methods**. As the basis for OpenGEM26 dataset, we selected molecules containing up to ten heavy (non-hydrogen) atoms from the GDB-13 dataset[32], which enumerates the chemical space of small organic molecules composed of H, C, N, O, S, and Cl. For each molecular formula, the GDB-13 dataset only includes chemical connectivity information encoded as SMILES strings. Based on these SMILES strings, initial 3D structures were obtained by using the MMFF94 force field[53] via the gen3d option in Open Babel[54].

Afterwards, we conducted first principles calculations using Q-Chem software[55]. In all calculation processes, the accuracy is within ωB97X -D/Def2-SVP for first step and then use Def2-TZVP to get more accurate optimized structures. ωB97X-D functional includes empirical atom–atom dispersion corrections, so it can capture the weak interactions in molecules. Our dataset also records the process of structural optimization together, including the forces of each atom in each step, which can be used to train machine learning models later.

**Diversity comparison.** Here, we compared the dataset with the most widely used QM9 dataset, as shown in Figure 1(a), visualizing the energy in the dataset. The horizontal axis represents the atomization energy of the molecule, measured in eV, while the vertical axis represents the logarithmic coordinate of quantities. The higher atomization energy region corresponds to non-equilibrium configurations. As indicated by the energy distribution, OpenGEM26 contains a substantial number of molecular structures far from equilibrium, particularly in the region with atomization energies above zero.

Furthermore, in order to compare the atomic local environments of the two datasets, we first conducted statistical analysis on the bond lengths between two neighbor atoms, as shown in Figure 1(b)-(f). The vertical axis represents the bond length in Angstrom units, and the horizontal axis represents the logarithm of the quantity. OpenGEM26 and QM9 are located on both sides of the image. Firstly, from the perspective of bond length distribution range, OpenGEM26 can basically cover the range of QM9. However, due to the inclusion of many non-equilibrium structures, as expected, there are still many conformational spatial regions that cannot be covered by only equilibrium state data. Furthermore, in terms of quantity, OpenGEM26 has also increased by an order of magnitude compared to QM9.

Afterwards, we further calculated the angle information between the three bonding atoms, as shown in Figure 1(g)-(i). In this type of polar coordinate graph, the value of r represents the average bond lengths between the central atom and the other two atoms, and $\theta$ is the angle value. This image reflects the distribution of local structures composed of more atoms in the conformational space of the dataset. The blue density plot is from OpenGEM26 dataset, and the yellow density plot is from QM9 dataset. It can be seen that the diversity of the configuration space covered by OpenGEM26 is greater than that of the QM9 dataset.

Figure 2 provides a more intuitive comprehensive comparison of the two datasets in the chemical space. (a) shows the t-Stochastic Neighbor Embedding[56] (t-SNE) of the two datasets and their differential distributions, indicating that the richness of the structures of OpenGEM26 is superior in the vast majority of cases. In(b)-(i), we use features of the environment of each atom, the activation values that form the first layer of the ANI-1x model, to compare the chemical and conformational space coverage for the QM9 and OpenGEM26. For visual comparison, we applied the parametric t-Distributed Stochastic Neighbor Embedding (pt-SNE) technique[57] which learns a neural network-based mapping from high-dimensional vectors to a 2-dimensional representation, while preserving both global and local structure of the data in high-dimensional space as much as possible. We used the complete OpenGEM26 to learn a pt-SNE mapping neural network from the machine learning model activation vectors to a 2D space. Then we apply this deep neural network to the whole QM9 dataset and a random subset of OpenGEM26 with the same number of atoms for each element. The colors correspond to the type and number of bonded neighbors as determined by the Open Babel[54] software library.

## MACHINE LEARNING

**Model.** The deep learning model we use is derived from GPTFF[58], which utilizes the architecture of graph neural networks(GNNs)[59–62]. For molecular structure, GNNs is used to represent its structure. It is natural to use nodes to represent atoms, and the connection between nodes to represent the bonding of atoms. In this way, each graph can represent a molecule, and can naturally meet the symmetry of rotation and translation[10,59,61]. For better machine learning training, we will project atoms into high-dimensional space according to their element types to form element embeddings. The local environment information of each atom, such as the distance from the nearest neighbor atom, will be mapped to the edge vector through a certain transformation.

In addition, model also has a module to calculate the three-body interaction between atoms, which is realized by connecting the nodes and edge vectors of the corresponding atoms. This enables the model to learn the interaction between different atomic nodes and further nodes and edge vectors, which significantly improves the prediction ability of the model.

In the process of model prediction and reasoning, GNNs predicts the vector within the atomic cut-off radius into the form of atomic energy through the output block, and finally represents the sum of atomic energy as the total energy of the molecule. For the calculation of atomic forces, we use the method of automatic differentiation. The force on an atom can be obtained from the negative gradient of the total energy E with respect to the atomic coordinates, i.e., $\vec{F} = -\frac{\partial E}{\partial \vec{x}}$.

**Training results.** Using OpenGEM26 dataset to train the model, we get GPTFF-mol, a machine learning force field suitable for small molecules. In Figure 3(a), the parity plot compares predicted energies with DFT reference values, showing excellent agreement, mean absolute error (MAE) 16 meV per molecule, that is, 0.82 meV/atom. It can be seen that due to the large amount of data and wide coverage of conformational space, we get high accuracy of energy prediction.

Furthermore, as OpenGEM26 covers force data, it is possible to train models to predict the force acting on each atom in the molecule, thereby obtaining the molecular potential for quickly optimizing the structure of the molecule with DFT accuracy. Figure 3(b) shows the results of this model in predicting force on the test set, and compares it with the opensource ANI-2x model in the same dataset (Figure 3(c)). It can be seen that the prediction of force is close to the most accurate model currently reported, and is superior to the publicly available ANI-2x model. In summary, this model can achieve molecular structure optimization with near

DFT accuracy under the time consumption of molecular force field, providing assistance for large-scale high-throughput screening.

## APPLICATION

**Butane rotation.** For butane, we investigate the variation of energy predicted by different models with respect to rotation angle after rotating atomic groups. In Figure 4(a), we rotated the top methyl group along its direction with the nearest neighboring C atom. At the beginning, it is the structure corresponding to the minimum energy value. The blue line represents the energy calculated by DFT, which is used as a reference value for comparison. In addition to using OpenGEM26 to train the GPTFF model, we also selected the open-source version of ANI-2x[63] and Equivariant Transformer (ET) models implemented in TorchMD-Net[64,65], trained on fundamentally different datasets (QM9, ANI-1). From this figure, we can see that GPTFF-mol has similar accuracy to ANI-2x, slightly better than ET_ani1.

The energy of n-butane is significantly influenced by the dihedral angle of its central C2-C3 bond due to the varying steric interactions between the two bulky methyl groups. This is because the substituents on C1 and C4 (three hydrogens) are much smaller than C2 and C3 (two hydrogens and one methyl group). The energy changes associated with this rotation are primarily due to torsional strain from methyl-methyl eclipsing interactions, which are considerably stronger than the hydrogen-hydrogen and hydrogen-methyl interactions[66]. Rotation around the central C2-C3 bond leads to various conformational isomers, each possessing a distinct energy. The torsional strain resulting from unfavorable interactions between substituents dictates these energy differences. For instance, the anti-conformation (180°) is the most stable due to minimized steric hindrance, while the gauche-conformations (±60°) are higher in energy. The eclipsed conformations (0°, ±120°) represent energy maxima due to severe steric repulsion[66,67]. Understanding this potential energy surface is crucial for predicting molecular behavior and will be further explored through the following computation.

In Figure 4(b), the atomic group on one side of the C2–C3 bond was rotated rigidly about the C2–C3 axis. The initial position is shown in the small picture. In this example, GPTFF-mol has similar prediction accuracy to ANI-2x and ET_ani1, and both are far better than ET_qm9. The performance of the ET model trained on the QM9 dataset is far inferior to other models, which may be due to the lack of non-equilibrium structured data in the QM9 dataset.

**Keto-enol Tautomerism.** Keto-enol tautomerism is a fundamental concept in organic chemistry, describing the interconversion between a keto form (containing a carbonyl group) and an enol form (containing a hydroxyl group directly attached to a carbon-carbon double bond). This dynamic equilibrium, often acid- or base-catalyzed, involves the migration of a hydrogen atom and a concomitant rearrangement of bonding electrons[68]. The relative stability of the keto versus enol tautomer is highly dependent on various factors, including substitution patterns, solvent effects, and the presence of intramolecular hydrogen bonding or aromaticity. In general, the keto form is generally more stable for simple aldehydes and ketones[69]. However, with the participation of solvent molecules (such as water molecules), the transition energy barrier between keto form and enol form can be significantly reduced[70–73]. Understanding this tautomeric equilibrium is crucial for predicting reaction pathways and product distributions in numerous organic transformations. The following computational investigation aims to explore the energetic landscape of molecules in vacuum or water, and compare the accuracy of different models in predicting the energy barriers.

Figure 5(a) is a schematic diagram of a reversible reaction in the presence of a water molecule, while (b)-(d) show the energy changes when the molecule without water molecule, binds to one water molecule, and binds to two water molecules, respectively. The horizontal coordinate is the intrinsic reaction coordinate (IRC), which traces the minimum-energy path downhill from a transition state toward the connected reactant and product valleys[74]. For ease of comparison, the minimum-energy point on the DFT energy profile was taken as the zero-energy reference. The energies predicted by the other models were shifted accordingly, such that their energies at the corresponding IRC step were also set to zero. Transition-state searches and IRC calculations were performed with Q-Chem software. The left side is the keto form and the right side is the enol form. Afterwards, different molecular potential functions are used to calculate the energy of each structure along the reaction path, and their respective energy change curves are obtained. In Figure 5(c) and (d), the energy barriers of ET_ani1 are too large, so they are not shown in the figures. It can be seen that OpenGEM26 contains many molecular structures far from equilibrium, so our trained model GPTFF-mol can better predict the energy under non-equilibrium conditions such as saddle point on the potential energy surface and its vicinity. This is also conducive to further use of the model for molecular structure optimization and other tasks.

## CONCLUSION

In this work, a comprehensive molecular dataset containing abundant equilibrium and non-equilibrium structures is constructed based on the GDB-13 database via high-precision

DFT calculations. Covering molecules composed of H, C, N, O, S and Cl with up to ten heavy atoms, this dataset features full structural optimization trajectories and extensive non-equilibrium conformations, which greatly expands the chemical and conformational space compared with conventional benchmarks such as QM9. Built upon the GPTFF graph neural network framework, the GPTFF-mol molecular potential is trained using the newly established dataset.

The developed potential achieves accuracy approaching DFT calculations while retaining the high computational efficiency of classical force fields. Validated by butane torsional rotation and keto-enol tautomerization simulations, the model demonstrates reliable performance in describing molecular steric effects, torsional strain and reaction energy barriers, even for distorted geometries near potential energy saddle points.

This work provides a high-quality dataset and a robust machine learning potential for simulating organic molecules containing sulfur and chlorine. The proposed framework facilitates large-scale high-throughput molecular screening and molecular dynamics simulations, and offers a promising route to explore complex organic systems and accelerate the discovery of novel functional molecular materials.

**Data/Code availability:**

The OpenGEM26 dataset will be openly released after peer review. GPTFF-mol model, training scripts, and weights are available at https://github.com/atomly-materials-research-lab/gptff/tree/main/pretrained/molecular.

**Author Contributions**:

S.M. and M.L. conceived and supervised the project. Y.H. and F.X. performed dataset construction, first-principles calculations, and model training. J.Z. contributed to the conformational-diversity analysis and t-SNE embeddings. Y.H. drafted the original manuscript with input from T.L. All authors discussed the results and approved the final version. The authors declare no competing financial or personal interests that could have appeared to influence the work reported in this paper.

**ACKNOWLEDGMENTS**:

This research is supported by National Key R&D Program of China (Grant No. 2024YFF0508500), Ministry of Science and Technology (Grant Nos. 2021YFA1400201 and 2021YFA1400503), National Natural Science Foundation of China (Grant Nos. 12025407 and 12450401), and Chinese Academy of Sciences (Grant No. YSBR047). The computational resources are provided by the Data-Driven Computational Materials Discovery Platform at Songshan Lake laboratory.

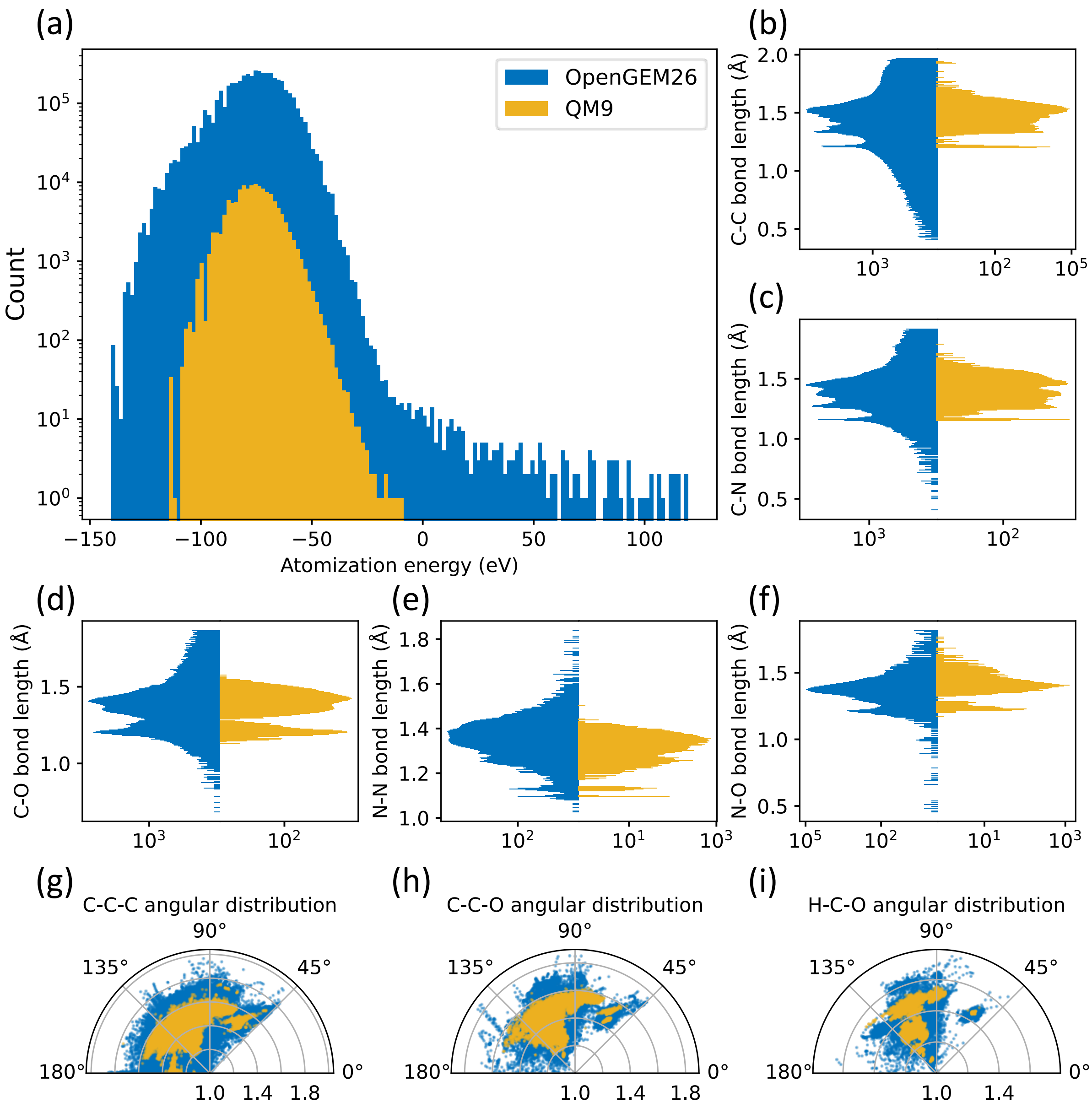

**Figure 1.** Properties distribution of QM9 and OpenGEM26 dataset. (a) Atomization energy, the vertical axis represents the logarithm of the structure number in two datasets. (b)-(f) show the bond lengths distribution images for C-C, C-N, C-O, N-N and N-O bonds, respectively. The x-axis represents the logarithm of the number. The y-axis represents the bond length of this type of bond. (g)-(i) Display the bond angle distribution images of C-C-C, C-C-O, and H-C-O bonds respectively. The angle of polar coordinates represents the bond angle of the type of triatomic bond, while the radius represents the average bond length between the central atom and the other two atoms. The blue dots in figures represent the distribution of the OpenGEM26 dataset, while the yellow dots represent the distribution of QM9 dataset.

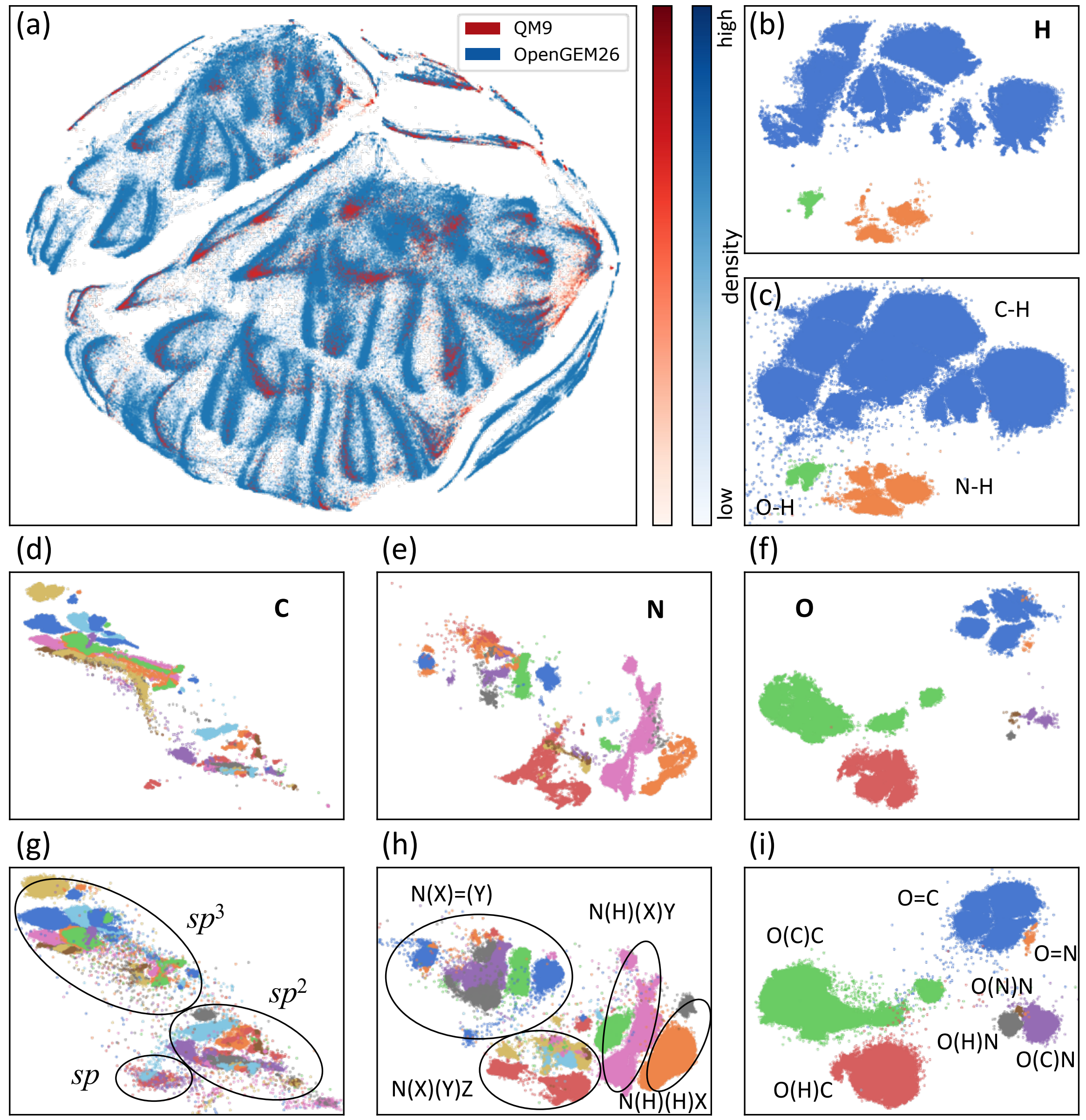

**Figure 2.** (a). The two-dimensional t-SNE embedding representation of the QM9 dataset and OpenGEM26, where the colors of the points are labeled according to the data density distribution, with red representing the QM9 dataset and blue representing OpenGEM26. (b-i) 2D parametric t-SNE embeddings. These embeddings are for the 1st layer of activations of the ANI-1x model for the complete QM9 dataset ((b) and (d-f)) and random subsets of OpenGEM26 dataset((c) and (g-i)). The same number of atoms are compared for each element. The different colors correspond to the number and type of bonded neighbors.

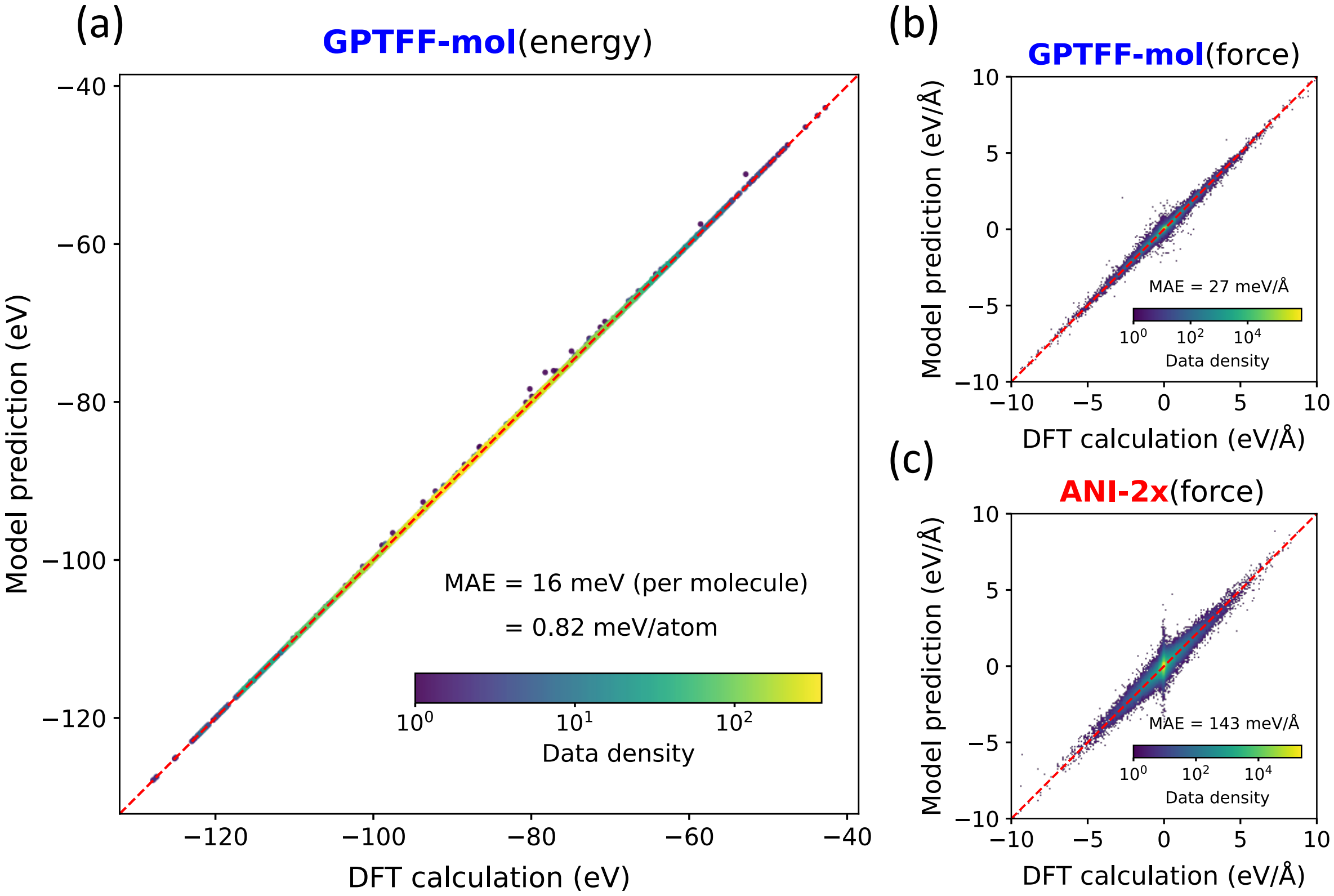

**Figure 3.** (a) The evaluation of MAE results of GPTFF-mol on atomization energy regression targets in the test dataset. (b) The evaluation of MAE results of GPTFF-mol on force regression targets in the test dataset. (c) The ANI-2x model performance on force regression targets in the same test dataset.

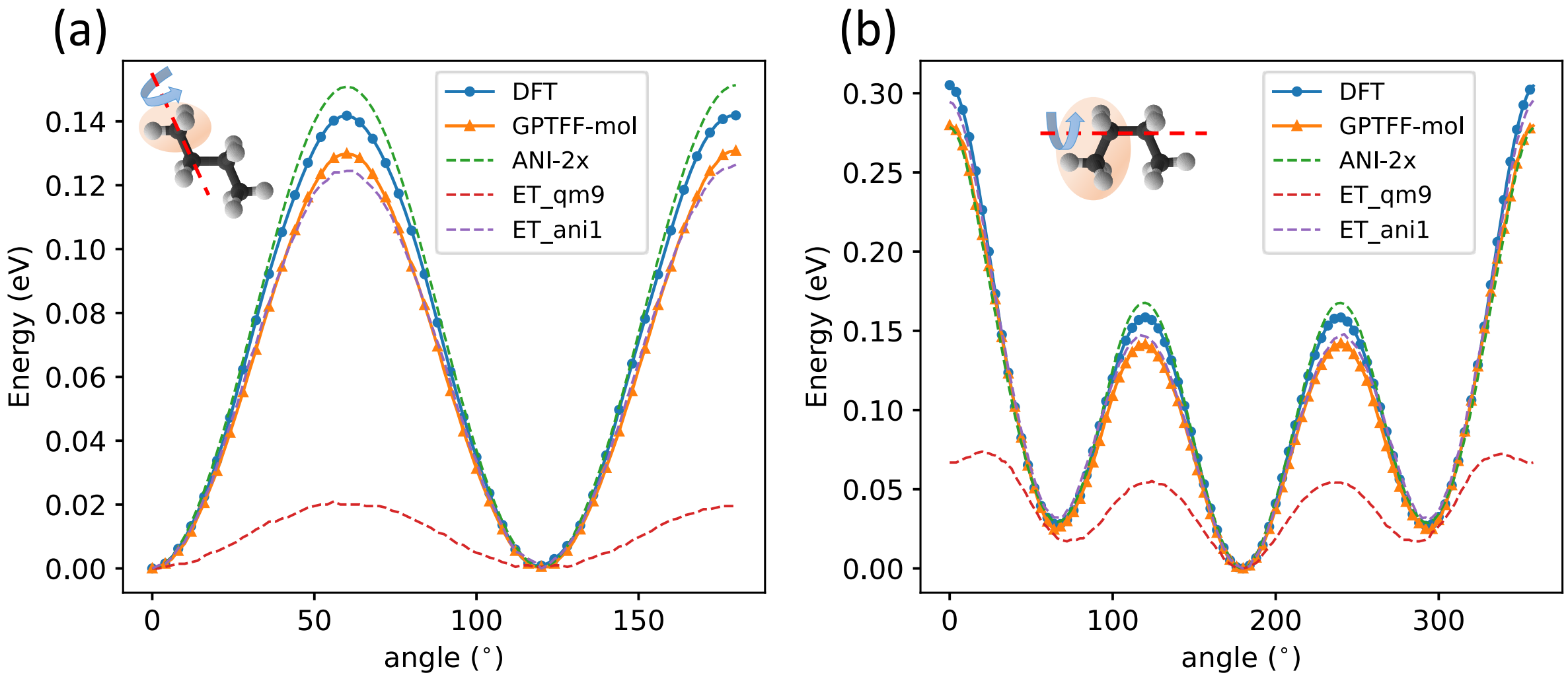

**Figure 4.** Butane rotational energy curves around two different bonds. (a) Rotation about the C1–C2 bond. (b) Rotation about the C2–C3 bond, where steric repulsion between methyl groups produces a higher barrier.

Density functional theory (DFT) results (blue lines) are compared with predictions from GPTFF-mol, ANI-2x, ET_qm9, and ET_ani1. Insets depict the corresponding rotation axes.

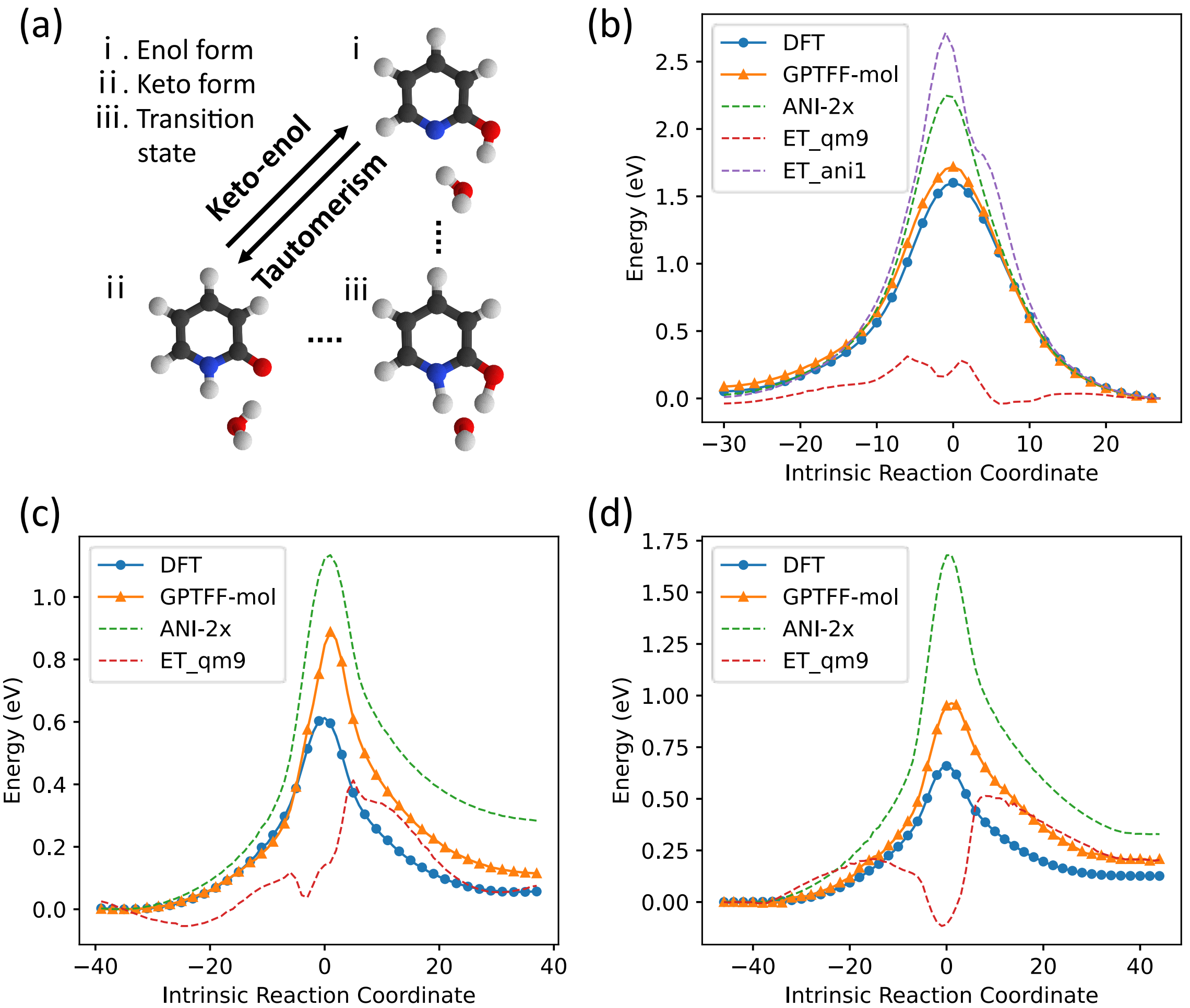

**Figure 5.** Keto–enol tautomerization of 2-hydroxypyridine and 2-pyridone. (a) Schematic representation of the interconversion between the two tautomers. (b–d) Energy vs. intrinsic reaction coordinate (IRC) plot for tautomerization in the absence of water, and with one and two water molecules, respectively. The left side is the keto form and the right side is the enol form. Density functional theory (DFT) results (blue lines) are compared with predictions from GPTFF-mol, ANI-2x, ET_qm9, and ET_ani1 (panel b only).